\documentclass[lettersize, journal]{IEEEtran}
\usepackage{amsmath,amsfonts,amssymb}
\usepackage{bm}
\usepackage{algorithmic}
\usepackage{algorithm}
\usepackage[caption=false,font=scriptsize,labelfont=sf,textfont=sf]{subfig}
\usepackage{url}
\usepackage{verbatim}
\usepackage{graphicx}
\usepackage{cite}
\usepackage{xcolor}
\begin{document}
\title{Flexible Intelligent Metasurface for Reconfiguring Radio Environments}
\author{Hanwen Hu, Jiancheng An,~\IEEEmembership{Member,~IEEE}, Lu Gan,~\IEEEmembership{Member,~IEEE}, and Naofal Al-Dhahir, \IEEEmembership{Fellow, IEEE}
\thanks{Copyright (c) 20xx IEEE. Personal use of this material is permitted. 
However, permission to use this material for any other purposes must be obtained 
from the IEEE by sending a request to pubs-permissions@ieee.org.}
\thanks{This work is supported by National Natural Science Foundation of China 62471096. The work of N. Al-Dhahir was supported by Erik Jonsson Distinguished 
Professorship at UT-Dallas.}
\thanks{H. Hu and L. Gan are with the School of Information and Communication Engineering, University of Electronic Science and Technology of China (UESTC), Chengdu, Sichuan, 611731, China. L. Gan is also with the Yibin Institute of UESTC, Yibin 644000, China (e-mail: hanwen\_hu\_uestc@outlook.com; ganlu@uestc.edu.cn). J. An is with the School of Electrical and Electronics Engineering, Nanyang Technological University, Singapore 639798 (e-mail: jiancheng\_an@163.com). N. Al-Dhahir is with the Department of Electrical and Computer Engineering, The University of Texas at Dallas, Richardson, TX 75080 USA (e-mail: aldhahir@utdallas.edu).}\vspace{-0.5cm}}

\maketitle
\begin{abstract}
\textbf{Flexible intelligent metasurface (FIM) technology holds immense potential for increasing the spectral efficiency and energy efficiency of wireless networks. In contrast to traditional rigid reconfigurable intelligent surfaces (RIS), an FIM consists of an array of elements, each capable of independently tuning electromagnetic signals, while flexibly adjusting its position along the direction perpendicular to the surface. In contrast to traditional rigid metasurfaces, FIM is capable of morphing its surface shape to attain better channel conditions. In this paper, we investigate the single-input single-output (SISO) and multiple-input single-output (MISO) communication systems aided by a transmissive FIM. In the SISO scenario, we jointly optimize the FIM phase shift matrix and surface shape to maximize the end-to-end channel gain. First, we derive the optimal phase-shift matrix for each tentative FIM surface shape to decompose the high-dimensional non-convex optimization problem into multiple one-dimensional subproblems. Then, we utilize the particle swarm optimization (PSO) algorithm and the multi-interval gradient descent (MIGD) method for updating the FIM's surface shape to maximize the channel gain. In the MISO scenario, we jointly optimize the transmit beamforming, the FIM surface shape, and the phase shift matrix to maximize the channel gain. To tackle this complex problem with multiple highly coupled variables, an efficient alternating optimization algorithm is proposed. Simulation results demonstrate that FIM significantly improves channel gain compared to traditional RIS and exhibits good adaptability to multipath channels.}
\end{abstract}


\IEEEpeerreviewmaketitle
\section{Introduction}
\IEEEPARstart{R}{econfigurable} intelligent surface (RIS) has emerged as a promising technology to transform wireless communication systems \cite{ref1,ref2,ref3,ref4,ref20}. It enjoys strong compatibility with existing wireless communication technologies, including massive multiple-input multiple-output (MIMO) systems and millimeter-wave (mmWave) communications\cite{ref5,ref6}. Through intelligent manipulation of the response of electromagnetic waves, RIS can dynamically reshape wireless propagation environments. More recently, various innovative metasurface architectures, such as dynamic metasurface antennas (DMA) and stacked intelligent metasurfaces (SIM), have been developed to show remarkable progress in terms of tuning capability\cite{ref7,ref8}.

Nonetheless, conventional RIS generally relies on a rigid substrate, which enhances the channel gain by adjusting the phase shift to superimpose multiple reflected and/or transmitted channels. However, the performance brought by the rigid RIS is limited by severe fading caused by practical multipath effects. To address this challenge, an advanced flexible intelligent metasurface (FIM) has been proposed recently, which can further improve the performance of RISs\footnote{\url{https://www.eurekalert.org/multimedia/950133} provides a video demonstrating the real-time morphing ability of an FIM.} \cite{ref9}. Specifically, FIMs can intelligently morph their surface shapes to adapt the steering vectors for constructively combining multipath signal components, thereby enhancing the received signal quality via each FIM element. This capability is particularly critical for millimeter-wave and terahertz communications, where the channel coherence distance is typically short. Currently, various FIM prototypes have been developed. In \cite{ref13} and \cite{ref14}, the authors implemented FIMs controlled by distributed Lorentz force. In \cite{ref15}, a bilayer FIM structure achieved controllable surface shape deformation. Recent advancements in FIMs have enabled their deployment as multifunctional interfaces for diverse applications including flexible robotics, intelligent embedded skins, and wearable sensors. In contrast to rigid RISs, FIMs show significant potential for enhancing smart radio environments by adjusting both their surface shape and phase shifts\cite{ref19,ref17,ref16}, thus providing additional design degree of freedom (DoF) to further improve the channel gain\cite{ref21}.

To analyze the performance gain of FIMs, we investigate the single-input single-output (SISO) and multiple-input single-output (MISO) systems aided by a transmissive FIM in a multipath fading channel. In the SISO scenario, we first derive the optimal phase shifts for each tentative FIM surface shape. Hence, the high-dimensional non-convex optimization problem is decomposed into multiple subproblems, each with respect to the deformation of a single element. The particle swarm optimization (PSO) algorithm\cite{ref11} and the multi-interval gradient descent (MIGD) method are used to update the FIM's surface shape to maximize the channel gain. In the MISO scenario, the optimization problem becomes more complex, since the transmit beamforming is also considered. To address this issue, an efficient alternating optimization algorithm is proposed. Specifically, for a fixed transmit beamforming vector, the problem is simplified to the SISO model, and for a given surface shape and FIM phase shift matrix, the maximum ratio transmit beamformer can be readily obtained. By iteratively performing these two steps, the channel gain is progressively improved.

\textit{Notations}: Vectors and matrices are denoted by bold-face lower-case and upper-case letters. The $n$-th element of vector $\bm{a}$ is $a_n$. The element in the $n$-th row and $m$-th column of matrix $\bm{A}$ is $A_{n,m}$. $\mathbb{C}^{x \times y}$ and $\mathbb{R}^{x \times y}$ denote the space of $x \times y$ complex-valued and real-value matrices. $(\cdot)^T$, $(\cdot)^H$, $|\cdot |$, and $||\cdot||$ denote the transpose, the conjugate transpose, the absolute value, and the Euclidean norm, respectively. For a complex vector $\bm{x}$, $\text{arg}(\bm{x})$ denotes a vector with each element being the phase of the corresponding element in $\bm{x}$, and $\text{diag}(\bm{x})$ denotes a diagonal matrix with each diagonal element being the corresponding element in $\bm{x}$. We use $\otimes$ and $\odot$ to represent the Kronecker product and Hadamard product, respectively. $\mathcal{CN}(\bm{a}, \bm{C})$ stands for the circularly-symmetric complex Gaussian (CSCG) distribution with mean vector $\bm{a}$ and covariance matrix $\bm{C}$.

\section{FIM-aided SISO Communications}
\subsection{SISO System Model and Problem Formulation}
\begin{figure}[t] 
\centering %
\includegraphics[width=0.45\textwidth]{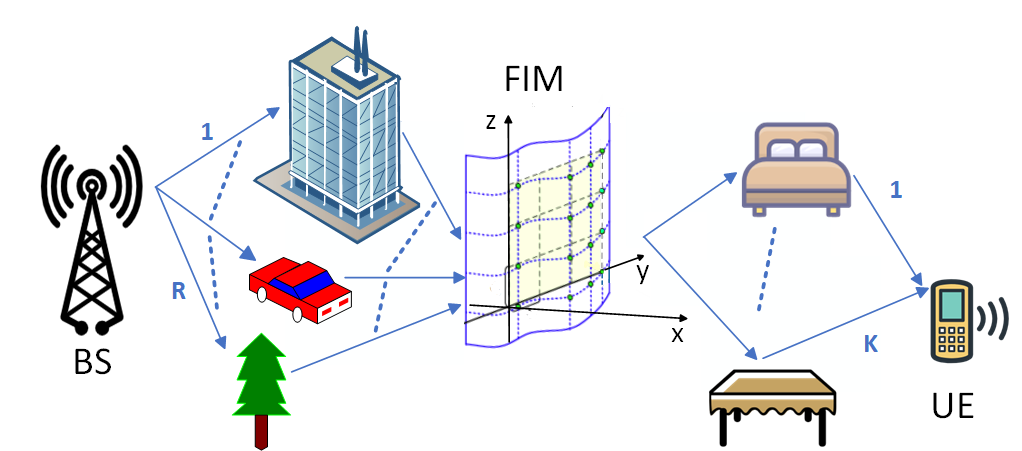}
\caption{An FIM-aided communication system.} 
\label{fig_1}
\vspace{-0.5cm}
\end{figure}

As shown in Fig. \ref{fig_1}, we consider a downlink communication link from the base station (BS) to a single user (UE), aided by a transmissive FIM composed of $N$ elements, with $N_{y}$ elements along the $y$-axis and $N_{z}$ elements along the $z$-axis. In contrast to a conventional RIS, the FIM surface shape can be adjusted by controlling the position of each element. Let \textit{$d_{n}$} represent the deformation distance of the $n$-th element with respect to the unmorphed substrate. As a result, the surface shape of the FIM is characterized by $\bm{d}=\{d_{1},d_{2},\ldots,d_{N}\}\in\mathbb{R}^{N \times 1}$.

The unmorphed FIM is modeled by a uniform planar array (UPA) with half-wavelength spacing. For a given azimuth angle $\theta$ and an elevation angle $\phi$, the steering vector $\bm{a}_{upa}(\theta, \phi)\in \mathbb{C}^{N \times 1}$ is represented as
\begin{equation}
\bm{a}_{upa}(\theta, \phi) = \bm{a}_y(\theta, \phi) \otimes\ \bm{a}_z(\phi),
\end{equation}
where $\bm{a}_y(\theta, \phi)\in \mathbb{C}^{N_y \times 1}$, \text{and} $\bm{a}_z(\phi) \in \mathbb{C}^{N_z \times 1}$ are defined by
\begin{equation}
\bm{a}_y(\theta, \phi) = 
\begin{bmatrix} 
1,e^{j\pi\sin{\theta}\cos{\phi}},\ldots,e^{j\pi(N_y-1)\sin{\theta}\cos{\phi}} 
\end{bmatrix}^T,
\end{equation}
\begin{equation}
\bm{a}_z(\phi) = 
\begin{bmatrix}
1,e^{j\pi\sin{\phi}},\ldots,e^{j\pi(N_z-1)\sin{\phi}} 
\end{bmatrix}^T.
\end{equation}

Furthermore, the additional response $\bm{a_}{d}(\theta,\phi,\bm{d})$ $\in \mathbb{C}^{N \times 1}$ caused by the surface-shape morphing is represented as
\begin{equation}
\bm{a_}{d}(\theta,\phi,\bm{d})=e^{j\kappa\bm{d}\cos{\theta}\cos{\phi}},
\end{equation}
where $\kappa=2\pi/{\lambda}$ represents the wavenumber, while $\lambda$ is the carrier wavelength. Consequently, the effective steering vector $\bm{a}$($\theta,\phi, \bm{d}$) is derived as 
\begin{equation}
 \bm{a}(\theta,\phi,\bm{d})=
 \bm{a}_{upa}(\theta, \phi) \odot \bm{a}_{d}(\theta,\phi,\bm{d}).
\end{equation}

Furthermore, let $R$ represent the number of paths between the BS and the FIM, with $\alpha_r$ representing the gain of the $r$-th path. Similarly, we assume $K$ paths between the FIM and the UE, with $\beta_k$ representing the gain of the $k$-th path. Moreover, let $\theta_r^I,\phi_r^I$ and $\theta_k^O,\phi_k^O$ represent the angles of arrival (AoA) and angles of departure (AoD) of the $r$-th BS-FIM path and the $k$-th FIM-UE path, respectively. In this paper, we aim to characterize the performance gain brought by the FIM and assume that the channel state information (CSI) is perfectly known.

Given these channel parameters, the baseband equivalent channel from the BS to the FIM and that from the UE to the FIM, denoted by $\bm{g}(\bm{d})$ $\in \mathbb{C}^{N \times 1}$ and $\bm{h}(\bm{d})$ $\in \mathbb{C}^{N \times 1}$ respectively, are given by
\begin{equation}
 \bm{g}(\bm{d}) = \sum_{r=1}^{R} \alpha_r \bm{a}(\theta_{r}^I,\phi_{r}^I,\bm{d}),
 \end{equation}
\begin{equation}
 \bm{h}(\bm{d}) = \sum_{k=1}^{K} \beta_k \bm{a}(\theta_{k}^O,\phi_{k}^O,-\bm{d}).
\end{equation}

Note that opposite surface shapes are observed on both sides of the FIM. Moreover, let diagonal matrix $\bm{S} = \text{diag}( e^{j\varphi_1}, \ldots, e^{j\varphi_N})$ denote the FIM phase shift matrix, where $\varphi_n \in [0, 2\pi)$ denotes the phase shift of the $n$-th FIM element. Let $x$ represent the signal transmitted by the BS. As a result, the signal $y$ received at the UE is given by
\begin{equation}
 y = \bm{h}^H(\bm{d}) \bm{S} \bm{g(d)}x + n,
\end{equation}
where $n\sim \mathcal{CN}(0, \sigma^2)$ denotes the additive white Gaussian noise (AWGN). Therefore, the SISO channel gain $\Gamma_{\text{SISO}}$ is given by 
\begin{equation}
 \Gamma_{\text{SISO}} = |\bm{h}^H(\bm{d}) \bm{S}\bm{g(d)} |^2.
\end{equation}

In this paper, we aim to optimize $\bm{d}$ and $\bm{S}$ to maximize the end-to-end channel gain, subject to constraints of the phase shifts and the maximum surface-shape morphing range $d_{\text{max}}$. As a result, the optimization problem in the SISO scenario is formulated as
\begin{align}
\text{(P1)} :\quad& \max_ {\bm{S} ,\bm{d}} \quad|\bm{h}^H(\bm{d}) \bm{S} \bm{g(d)}|^2 \tag{10a}\\
&\text{s.t.} \quad \quad |d_n| \leq d_{\text{max}}, \tag{10b}\\
&\quad \quad \quad 0 \leq \varphi_n \leq 2\pi, \quad \forall n = 1, \ldots, N.\tag{10c}
\end{align}

\subsection{Joint 3D Surface-shape Morphing and Phase Shift Configuration}
In this subsection, we first tackle the coupling between optimization variables $\bm{S}$ and $\bm{d}$ in problem (P1). Then, the PSO and the MIGD algorithm are utilized to solve the simplified problem.
\begin{figure*}[!t]
\normalsize
\setcounter{equation}{13}
\begin{equation}
\label{eqn_dbl_z}
 z_n(d_n)
 =\frac{1}{2}\sum_{r=1}^{R} \sum_{r'=1}^{R}\sum_{k=1}^{K} \sum_{k'=1}^{K} \alpha_r\alpha_{r'}\beta_k\beta_{k'}
 \{\cos[f_{r,r'}(d_n)+t_{k,k'}(d_n)]+\cos[f_{r,r'}(d_n)-t_{k,k'}(d_n)]\}.
\end{equation}
\begin{equation}
\label{eqn_dbl_x}
 f_{r,r'}(d_n)=\frac{2\pi d_n}{\lambda}(\cos{\phi_r^I}\cos{\theta_r^I}-\cos{\phi_{r'}^I}\cos{\theta_{r'}^I})+\pi n_y(\sin{\phi_r^I}-\sin{\phi_{r'}^I})+\pi n_z(\sin{\theta_r^I}\cos{\phi_r^I}-\sin{\theta_{r'}^I}\cos{\phi_{r'}^I}).
\end{equation}
\begin{equation}
\label{eqn_dbl_y}
 t_{k,k'}(d_n)=\frac{-2\pi d_n}{\lambda}(\cos{\phi_k^O}\cos{\theta_k^O}-\cos{\phi_{k'}^O}\cos{\theta_{k'}^O})
 +\pi n_y(\sin{\phi_k^O}-\sin{\phi_{k'}^O})
 +\pi n_z(\sin{\theta_k^O}\cos{\phi_k^O}-\sin{\theta_{k'}^O}\cos{\phi_{k'}^O}).
\end{equation}
\hrulefill
\vspace{-0.5cm}
\end{figure*}

For brevity, we rewrite $\bm{h}^H(\bm{d}) \bm{S} \bm{g}(\bm{d}) = \bm{s}^H \bm{v}(\bm{d})$, where $\bm{s} = [e^{j\varphi_1}, \ldots, e^{j\varphi_N}]^H$ $\in \mathbb{C}^{N \times 1}$ and $\bm{v}(\bm{d}) = \text{diag}(\bm{h}^H(\bm{d})) \bm{g}(\bm{d})$ $\in \mathbb{C}^{N \times 1}$. Hence, problem (P1) is equivalent to
\begin{align}
\text{(P2)} : \quad & \max_ {\bm{s} ,
 \bm{d}} \quad |\bm{s}^H \bm{v}(\bm{d})|^2 \tag{11a}\\
&\text{s.t.} \quad \quad \text{(10b)}, \text{(10c)} .\tag{11b}\end{align}

It is not difficult to show that the optimal solution to problem (P2) has the form
\begin{equation}
 \bm{s^*} = e^{j( \arg( \bm{v}(\bm{d})))} = e^{j(\arg(\text{diag}(\bm{h}^H(\bm{d}))\bm{g}(\bm{d})))}.\tag{12}
\end{equation}

It is observed from (12) that the optimal solution for the phase shift vector $\bm{s}$ can be uniquely determined by the surface-shape morphing vector $\bm{d}$. By substituting (12) in (11a), problem (P2) can be further simplified to
\begin{align}
\text{(P3)} : \quad & \max_{\bm{d}} \quad {(\sum_{n=1}^{N} 
 | v_n(d_n)|)^2} \tag{13a}\\
&\text{s.t.} \quad \quad |d_n| \leq d_{\text{max}}, \quad \forall n = 1, \ldots, N.\tag{13b}
\end{align}

Note that the objective function in (13a) becomes the square of the sum of the channel magnitude associated with each element. Hence, problem (P3) can be solved by maximizing the channel gain associated with each FIM element. Specifically, we define $z_n(d_n)\triangleq|v_n(d_n)|^2$ and its detailed expression is shown in (14), where $n_y \in \{0,...., N_y-1\}$, and $n_z \in \{0,...., N_z-1\}$. Then, problem (P3) is equivalent to solving $N$ parallel optimization problems, yielding
\begin{align}
\text{(P4)} : \quad & \max_{d_n} \quad z_n(d_n) \tag{17a}\\
&\text{s.t.} \quad \quad |d_n| \leq d_{\text{max}}, \quad \forall n = 1, \ldots, N.\tag{17b}
\end{align}

Note that the channel gain associated with each element is the sum of $2K^2R^2$ cosine functions, which is a non-convex optimization problem. To solve problem (P4), we develop the following two effective algorithms.
\subsubsection{Particle Swarm Optimization}
A commonly used method to solve such a non-convex optimization problem is the PSO algorithm, which exhibits good convergence speed and robustness. Specifically, the PSO algorithm initializes several particles within the solution space (17b), where each particle is assigned an initial velocity and position, allowing it to freely explore the search space for finding local optima. During the search process, each particle dynamically adjusts its velocity and position by comparing its own historical best solution with the global best solution found by the entire swarm, thus facilitating information sharing and collaborative optimization. By doing so, the particle swarm gradually converges toward the global optimum through multiple iterations according to the following update equations
\begin{align}
q_{n,i}^{(t+1)} &= \omega q_{n,i}^{(t)}+c_1 r_1 (d_n^{{*(t)}} - d_{n,i}^{(t)}) + c_2 r_2 (d_{n,i}^{*(t)} - d_{n,i}^{(t)}),\tag{18}\\
d_{n,i}^{(t+1)} &= d_{n,i}^{(t)} + q_{n,i}^{(t+1)},\tag{19}
\end{align}
where $q_{n,i}^{(t)} $ and $d_{n,i}^{(t)}$ are the velocity and position of the $i$-th particle at the $t$-th iteration for the $n$-th element, respectively. $ d_{n,i}^{*(t)}$ and $d_n^{*(t)}$ are the best positions found by particle $i$ and the entire swarm, respectively. $\omega$ is the inertia weight, $ c_1 $ and $ c_2 $ are positive acceleration coefficients, while $r_1$ and $ r_2 $ are random numbers sampled from a uniform distribution in $[0,1]$.

After applying (18) and (19) several times, the position of each element will gradually converge to a global optimal solution. In this paper, the objective function (17a) represents the fitness function, and the positions of all particles are projected to the feasible set between $-d_{\text{max}}$ and $d_{\text{max}}$ at each iteration. Assuming the number of particles is $Q$ and the number of iterations is $T_{\text{PSO}}$, the computational complexity of the PSO algorithm used to solve Problem (P4) is
$\mathcal{O}(Q T_{\text{PSO}} K^2 R^2)$.

\subsubsection{Multi-interval Gradient Descent Algorithm (MIGD)}
Additionally, the gradient descent method can be utilized to solve problem (P4). Although the original objective function exhibits strong non-convexity globally, it remains convex in a relatively small range of $d_n$. Therefore, the MIGD algorithm is invoked to divide the whole search interval into multiple (denoted by $\delta$) small intervals to ensure local convexity. Thus, each sub-interval has a length of $2d_{\text{max}}/\delta$. Let $d_{n,j}^*$ $(j= 1, \ldots, \delta)$ denote the optimal solution by applying the MIGD in the $j$-th interval. As a result, the optimal position of the $n$-th element is determined by
\begin{equation}
 d_n^*=\max_{d_n \in \Delta} z_n(d_n),\tag{20}
\end{equation}
where $\Delta \overset{\triangle}{=} \{d_{n,1}^*,\ldots,d_{n,\delta}^*\}$. Due to the complexity of the derivative of the objective function $z_n(d_n)$ with respect to the deformation distance $d_n$, we approximate the gradient at each position through the first-order Taylor expansion, yielding
\begin{equation}
 z_n'(d_n) \approx (z_n(d_n+\xi)- z_n(d_n)) /\xi,\tag{21}
\end{equation}
where $\xi$ denotes a very small surface-shape morphing range.

Furthermore, the computational complexity MIGD algorithm is
$\mathcal{O}(\delta T_{\text{GD}} K^2 R^2)$,  where $T_{\text{GD}}$ represents the number of gradient descent iterations per interval.

Overall, since problem (P1) is decomposed into $N$ subproblems, the total complexity of the algorithm scales linearly with $N$, provided that the number of FIM elements remains within a reasonable range. \footnote{In fact, when $N$ becomes particularly large, the dimensionality of the surface shape vector $\bm{d}$ increases, which causes each iteration of the PSO and MIGD algorithms to require more computational effort. As a result, the overall complexity grows with a steeper slope with respect to $N$. However, when $N$ remains within a reasonable range, this effect can be ignored.}

\section{FIM-aided MISO Communications}
\subsection{MISO System Model and Problem Formulation}
In the MISO system, the BS is equipped with $M$ antennas, arranged in a uniform linear array (ULA) with half-wavelength spacing. Let $\gamma_r$ represent the AoD of the $r$-th path with respect to the BS. The steering vector $\bm{a}_{ula}(\gamma_r)\in \mathbb{C}^{M \times 1}$ is thus written as
\begin{equation}
\bm{a}_{ula}(\gamma_r) = 
\begin{bmatrix}
1,e^{j\pi\sin{\gamma_r}},\ldots,e^{j\pi(M-1)\sin{\gamma_r}} 
\end{bmatrix}^T.\tag{22}
\end{equation}

Based on (22), the baseband equivalent channel matrix $\bm{G} \in \mathbb{C}^{N \times M}$ from the BS to the FIM is given by
\begin{equation}
 \bm{G}(\bm{d}) = \sum_{r=1}^{R} \alpha_r \bm{a}(\theta_{r}^I,\phi_{r}^I,\bm{d})\bm{a}_{ula}^H(\gamma_r).\tag{23}
 \end{equation}

At the BS, we consider linear transmit precoding. Specifically, let $\bm{x} = \bm{w} s$ represent the complex baseband transmitted signal at the BS, where $s$ denotes the transmitted data and $\bm{w} \in \mathbb{C}^{M \times 1}$ is the corresponding beamforming vector. The signal $y$ received at the UE is then expressed as
\begin{equation}
 y = \bm{h}^H(\bm{d}) \bm{S} \bm{G(d)}\bm{w}s + n.\tag{24}
\end{equation}

Therefore, the MISO channel gain $\Gamma_{\text{MISO}}$ is given by 
\begin{equation}
 \Gamma_{\text{MISO}} = |\bm{h}^H(\bm{d}) \bm{S}\bm{G(d)} \bm{w} |^2.\tag{25}
\end{equation}

Similarly, we aim to optimize the transmit beamforming vector $\bm{w}$, the FIM surface shape morphing $\bm{d}$ and the phase shifts $\bm{S}$ to maximize the channel gain. Specifically, the optimization problem in the MISO scenario is given by 
\begin{align}
\text{(P5)} : \quad & \max_ {\bm{S} ,
 \bm{d},\bm{w}} \quad |\bm{h}^H(\bm{d}) \bm{S} \bm{G(d)}\bm{w}|^2 \tag{26a}\\
&\text{s.t.} \quad \quad \text{(10b)}, \text{(10c)},\tag{26b}\\
& \quad \quad \quad ||\bm{w}||^2 \leq P,\tag{26c}
\end{align}
where $P$ denotes the transmit power of the BS.
\subsection{Joint Transmit Beamforming, 3D Surface-shape Morphing and Phase Shift Configuration}
Note that problem (P5) becomes more complex due to the coupling of three groups of variables. To tackle this problem, we first rewrite the FIM phase shifts $\bm{S}$ as a function of $\bm{w}$ and $\bm{d}$, and then propose an effective alternating optimization algorithm to solve the simplified problem.

Specifically, we rewrite $\bm{h}^H(\bm{d}) \bm{S} \bm{G}(\bm{d})\bm{w} = \bm{s}^H \bm{u}(\bm{d},\bm{w})$, where $\bm{u}(\bm{d},\bm{w}) = \text{diag}(\bm{h}^H(\bm{d})) \bm{G}(\bm{d})\bm{w}$ $\in \mathbb{C}^{N \times 1}$. Similarly, the optimal phase shift vector $\bm{s}$ is given by
\begin{equation}
 \bm{s^*} = e^{j( \arg( \bm{u}(\bm{d},\bm{w})))}= e^{j(\arg(\text{diag}(\bm{h}^H(\bm{d}))\bm{G}(\bm{d})\bm{w}))}. \tag{27}
\end{equation}

Substituting (27) into (26) yields
\begin{align}
\text{(P6)} : \quad & \max_{\bm{d},\bm{w}} \quad {(\sum_{n=1}^{N} 
 | u_n(d_n,\bm{w})|)^2} \tag{28a}\\
&\text{s.t.} \quad \quad |d_n| \leq d_{\text{max}}, \quad \forall n = 1, \ldots, N.\tag{28b}\\
& \quad \quad \quad ||\bm{w}||^2 \leq P.\tag{28c}
\end{align}

Let $o_n(d_n,\bm{w})\triangleq|u_n(d_n,\bm{w})|^2$ represent the channel gain component associated with the $n$-th element. Thus, we have
\begin{equation}
o_n(d_n,\bm{w})=|\sum_{m=1}^{M} 
 w_m h_n^*(d_n) G_{n,m}(d_n)|^2.\tag{29}
\end{equation}

Similar to (14), the objective function $o_n(d_n,\bm{w})$ takes the form of the sum of multiple cosine functions, with variables $d_n$ and $\bm{w}$ being highly coupled. To address this issue, the alternating optimization strategy is utilized to solve problem (P6) by iteratively optimizing the following two sub-problems.

\subsubsection{Transmit Beamforming – MRT}
Given the FIM surface shape $\bm{d}$ and the phase shifts $\bm{s}$, the optimal transmit beamforming vector $\bm{w}$ can be obtained by employing the well-known maximal ratio transmission (MRT) beamforming scheme, i.e.,
\begin{equation}
 \bm{w}^*=\sqrt{P}\frac{(\bm{h}^H(\bm{d}) \bm{S} \bm{G}(\bm{d}))^H}{||\bm{h}^H(\bm{d}) \bm{S} \bm{G}(\bm{d})||}.\tag{30}
\end{equation}

\subsubsection{Surface-shape Morphing and Phase Shift Configuration - PSO}
Given the transmit beamforming vector $\bm{w}$, problem (P6) reduces to problem (P3) and can be readily solved by decoupling the problem into multiple one-dimensional search problems. For each sub-problem, the PSO algorithm is used to find the optimal position solution of each element, with $N$ optimized positions constituting the surface-shape morphing vector $\bm{d}$. Meanwhile, we can obtain the optimal phase shift configuration from (27).

The detailed steps of the alternating optimization method are summarized in Algorithm 1, and the following two facts guarantee its convergence. Firstly, for each subproblem, the optimal solution is obtained, thus ensuring that the objective value of problem (P6) does not diminish during the iterations. Secondly, the optimal value of (28a) is bounded by an upper bound due to the surface-shape constraint (28b) and power budget (28c). Therefore, the proposed algorithm is guaranteed to converge.

\begin{algorithm}[t]
\renewcommand{\algorithmicrequire}{\textbf{Input:}}
\renewcommand{\algorithmicensure}{\textbf{Output:}}
\renewcommand{\algorithmicrepeat}{\textbf{Repeat}}
\renewcommand{\algorithmicuntil}{\textbf{Until}}
\caption{The proposed alternating optimization algorithm for joint transmit beamforming, 3D surface-shape morphing and phase shift configuration}
\begin{algorithmic}[1]
\REQUIRE 
$\phi_r^I,\theta_r^I,\phi_k^O,\theta_k^O,\gamma_r,\alpha_r,\beta_k,N,R,K,M,\lambda,
d_{\text{max}}$.
\STATE Initialize the phase shifts $\bm{s^0}$ and the surface-shape morphing $\bm{d^0}$. Set the iteration counter to $i = 0$.
\REPEAT
 \STATE Optimize $\bm{w}$ with given $\bm{d}^{(i)}$ and $\bm{s}^{(i)}$. Denote the optimal transmit beamforming vector as $\bm{w}^{(i)}$.
 \STATE Optimize $\bm{d}$ and $\bm{s}$ with given $\bm{w}^{(i)}$. Denote the optimal surface-shape morphing vector and the optimal phase shift vector as $\bm{d}^{(i+1)}$ and $\bm{s}^{(i+1)}$, respectively. 
 \STATE Update the iteration counter by $i \leftarrow i + 1$.
\UNTIL{The increase of the channel gain is less than the preset threshold or $i$ exceeds the maximum tolerable number of iterations.} 
\ENSURE $\bm{w^*},\bm{d^*},\bm{s^*}$.
\end{algorithmic}
\end{algorithm}
\vspace{-0.5cm}

\begin{figure*}[!t]
\centering
\subfloat[$z_1$ versus $d_1$]{\includegraphics[width=4.4cm]{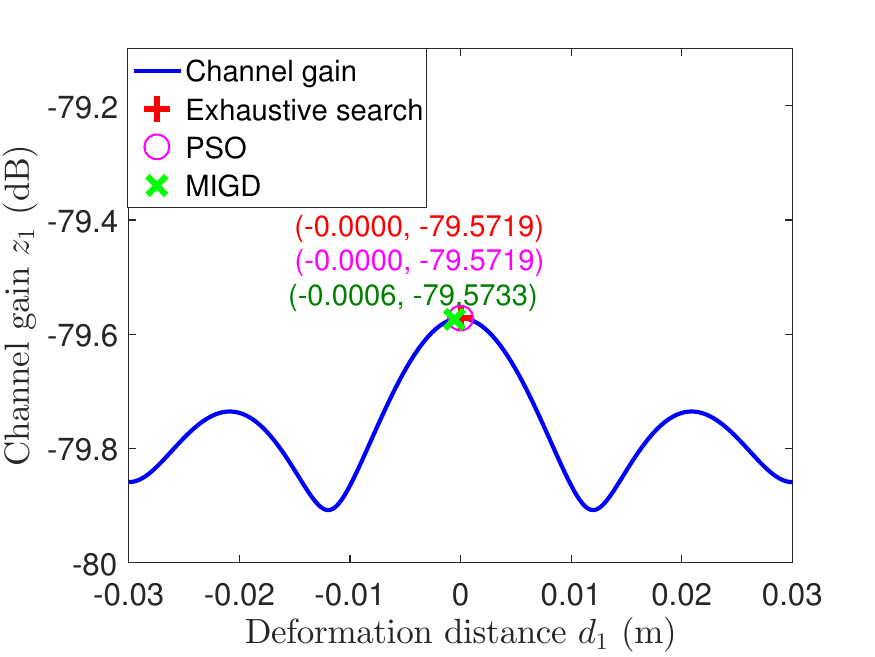}}
\subfloat[$z_2$ versus $d_2$]{\includegraphics[width=4.4cm]{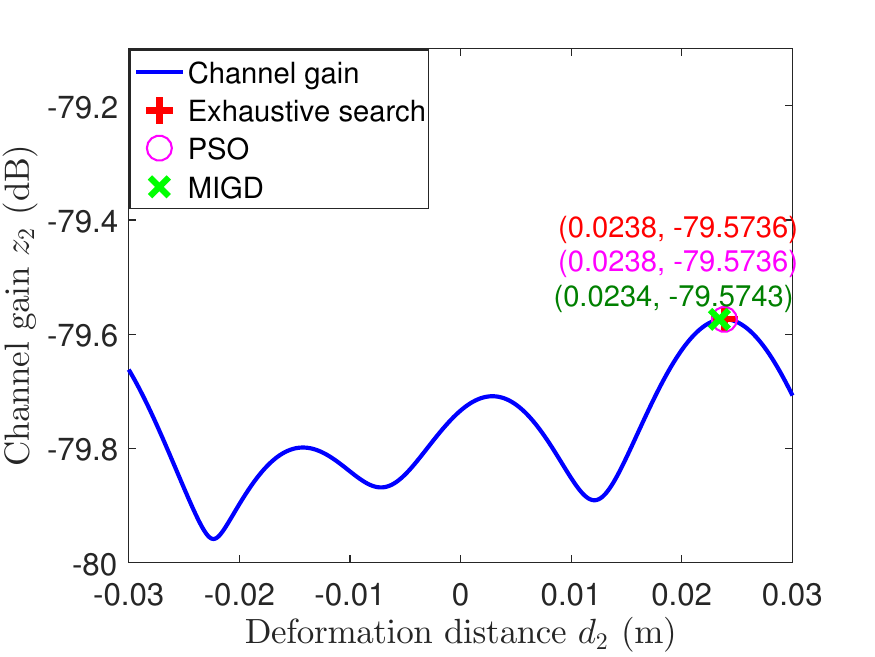}}
\subfloat[$z_3$ versus $d_3$]{\includegraphics[width=4.4cm]{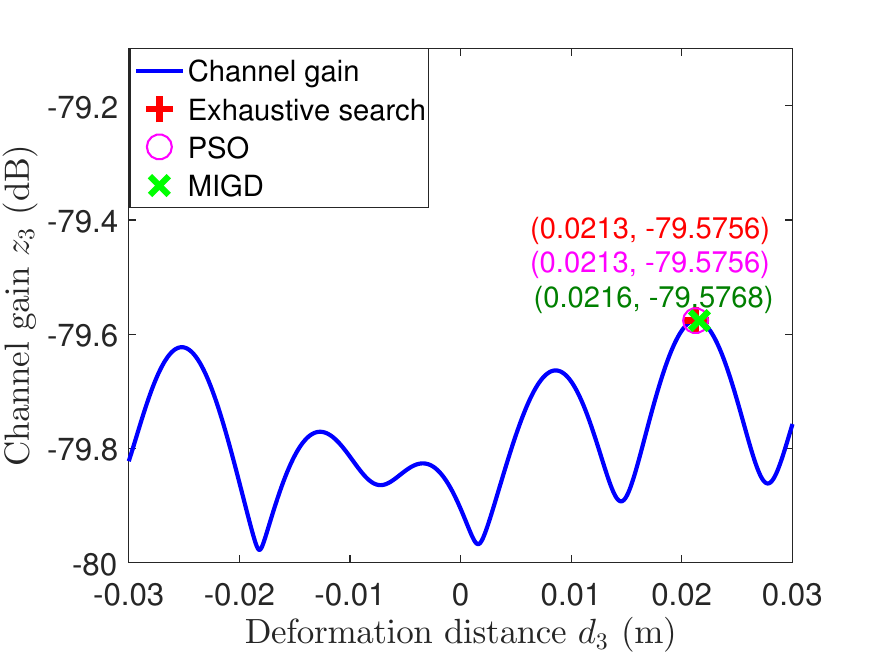}}
\subfloat[$z_4$ versus $d_4$]{\includegraphics[width=4.4cm]{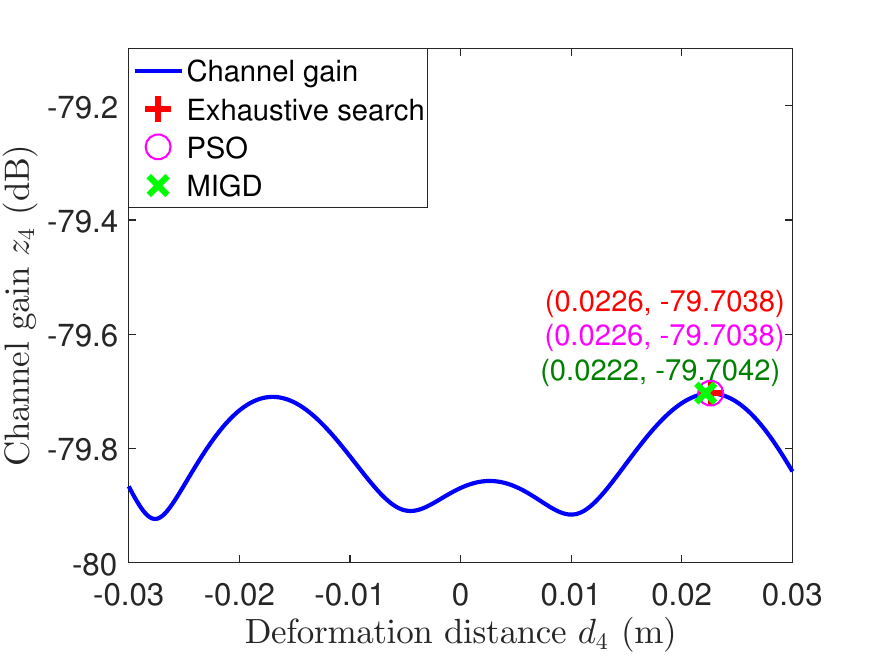}}
\caption{The channel gain $z_n$ versus the deformation distance of the $n$-th FIM's element (SISO).}
\label{fig_2}
\vspace{-0.5cm}
\end{figure*}
\begin{figure}[t]
\centering
\subfloat[]{\includegraphics[width=4.4cm]{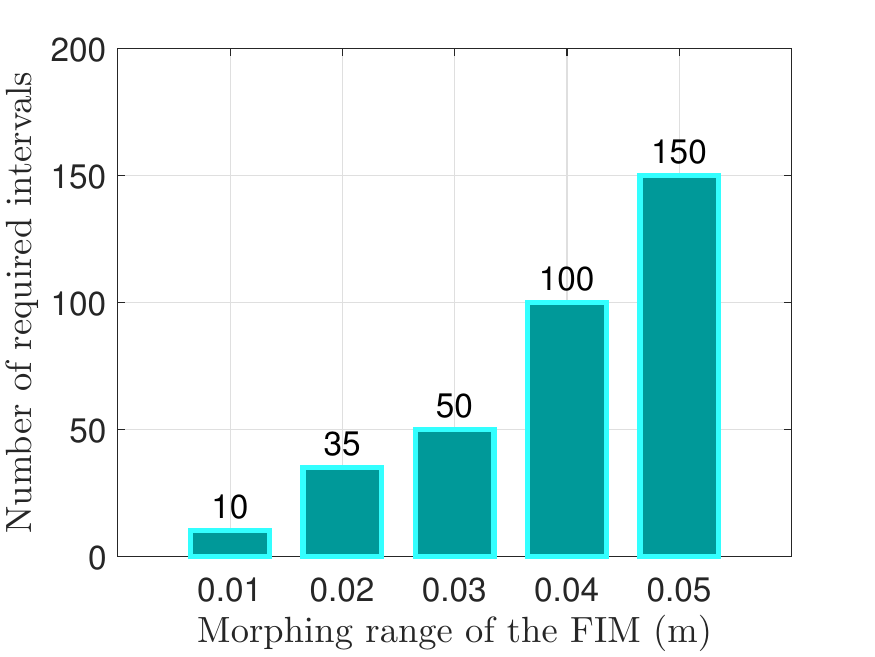}}
\subfloat[]{\includegraphics[width=4.4cm]{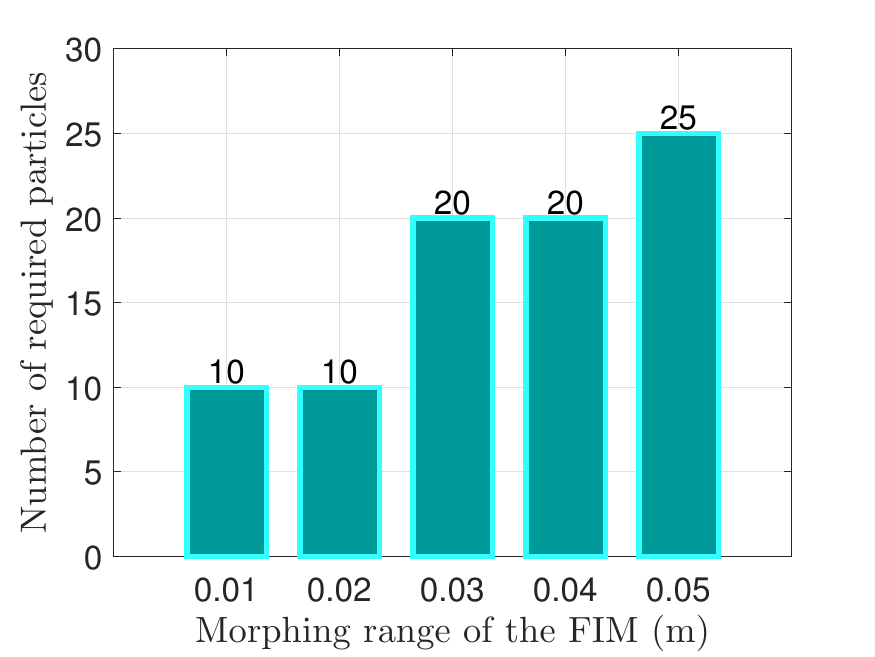}}
\caption{ (a) Number of intervals of MIGD versus morphing range (MIGD); (b) Number of particles of PSO versus morphing range (PSO).}
\vspace{-0.5cm}
\end{figure}
\section{Simulation Results}
In this section, we provide numerical results to validate the performance gain of FIM. Specifically, we consider a typical downlink communication system assisted by an FIM with $N=N_yN_z$ metasurfaces. We assume that the FIM surface shape morphing is continuous in its morphing range.\footnote{In practical scenarios, the deformation error of the FIM is much smaller than the wavelength and can be neglected. Moreover, the deformation response time (i.e., reconfigurability rate) of the FIM is on the order of milliseconds, which is comparable to the coherence time of the channel under typical mobile conditions, allowing it to adapt the wireless channel in real time.} For the purpose of exposition, we fix $N_z = 2$ and increase $N_y$ linearly. The BS-FIM distance is $ d_{\text{BF}} = 50$ m and the FIM-UE distance is $d_{\text{FU}} = 5$ m. Moreover, we assume $\alpha_r \sim \mathcal{CN}(0, \rho_R^2)$ and $\beta_k\sim \mathcal{CN}(0, \rho_K^2)$, while the distance-dependent path loss is determined by $\rho_R^2 = C_0 \left( d_{\text{BF}}/d_0 \right)^{-\ell_{\text{BF}}}$ and $\rho_K^2 = C_0 \left( d_{\text{FU}}/d_0 \right)^{-\ell_{\text{FU}}}$, where $C_0= -25$ dB is the path loss at the reference distance $d_0=1$ m. $\ell_{\text{BF}}$ and $\ell_{\text{FU}}$ are the path loss exponents, which are set to $\ell_{\text{BF}}=3.5$, $\ell_{\text{FU}}=2$. The AoA and AoD on both sides of the FIM $\phi_r, \phi_k, \theta_r,\theta_k$ and the transmit AoD with respect to the BS $\gamma_r$ are uniformly distributed over $[-\pi/2, \pi/2]$. The maximum number of iterations is set to 1000 and the stopping threshold is set as $\epsilon = 10^{-4}$. The noise power is $\sigma^2 = -80$ dBm and the wavelength is set as $\lambda = 0.01$ m. The transmit power is set as $P=15$ dBm. Other parameters of the PSO and MIGD algorithms are set as follows: $Q = 20$, $\omega = 0.8$, $c_1 = c_2 = 2$, $\delta= 50$. Unless otherwise specified, the maximum surface shape $d_{\text{max}}$ is set to $3\lambda$.

In Fig. \ref{fig_2}, we demonstrate the effectiveness of the proposed PSO and MIGD algorithms for maximizing the channel gain of FIM-aided SISO systems, where we set $N=4$ for illustration. The number of NLoS paths on both sides of the FIM is set to $K=R=3$. For comparison, the point coordinates of individual elements that maximize the channel gain are obtained by using exhaustive search and marked in Fig. \ref{fig_2}. From Fig. \ref{fig_2}, it can be observed that the proposed PSO and MIGD algorithm successfully match the exhaustive search results, while significantly reducing the complexity. The results obtained by MIGD, while exhibiting minor discrepancies compared to the other two methods, maintain a normalized deformation error margin strictly within 0.02\%. This marginal error originates from resolution limitations and can be effectively eliminated by employing more intervals and finer step sizes, albeit at the cost of increased computational complexity. As illustrated in Fig. 2, for the optimal surface shape, the signal components from multiple paths can be coherently combined at each individual element, thus enhancing the maximum channel gain. Furthermore, all elements attain their maximum channel gains at different displacement values, which further demonstrates that the effectiveness of FIM compared to conventional rigid RIS.

In Fig. 3, we present the hyperparameter of MIGD and PSO with respect to the morphing range. It can be observed that the MIGD algorithm is more sensitive to the morphing range, as a larger morphing range leads to a greater number of intervals. In contrast, PSO exhibits stronger robustness. Therefore, MIGD is more efficient in scenarios with a small morphing range and low precision requirements, whereas PSO performs better when higher precision or larger morphing ranges are involved.

\begin{figure}[!t]
    \centering
    \includegraphics[width=0.85\linewidth]{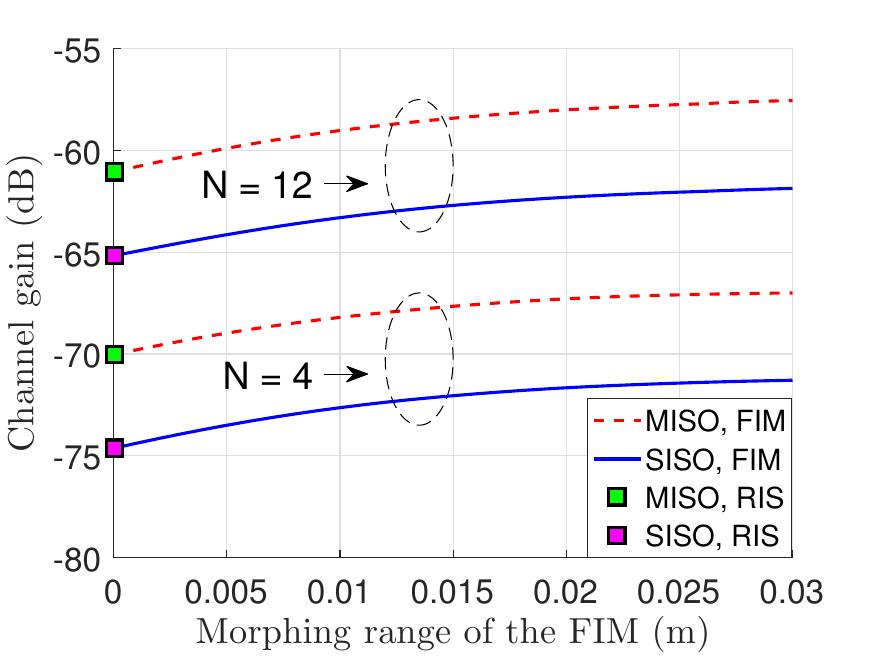}
    \caption{Channel gain versus the morphing range $d_{\text{max}}$.}
    \label{fig_3}
   \vspace{-0.3cm}
\end{figure}

Fig. \ref{fig_3} shows the channel gain versus the FIM morphing range, where we consider two setups: i) $N = 4$, and ii) $N = 12$. We also assume that there are $K=R=3$ propagation paths. For the MISO scenario, we consider $M=4$ transmit antennas at the BS, and the effective channel gain is calculated by dividing the channel gain in (26a) by $P$. It can be observed from Fig. \ref{fig_3} that as the maximum morphing range increases, the FIM significantly improves the channel gain compared to the traditional rigid RIS (i.e., $d_{\text{max}}=0$ m). By morphing its surface shape to enable constructive superposition of all propagation paths across the array, over $3$ dB performance gain is observed under all scenarios considered. Nevertheless, the channel gain exhibits a diminishing return as $d_{\text{max}}$ increases. This phenomenon arises from the fact that the objective function exhibits periodicity with respect to the morphing range since the FIM surface shape vector is embedded in the complex exponential term of the multipath steering vector. As a result, as the morphing range increases beyond a certain value, further increases in $d_{\text{max}}$ do not lead to additional performance gains. As such, the near-optimal surface shape can be obtained within a small morphing range, which demonstrates the advantages of FIM in practical applications.

\begin{figure}[!t]
    \centering
    \includegraphics[width=0.85\linewidth]{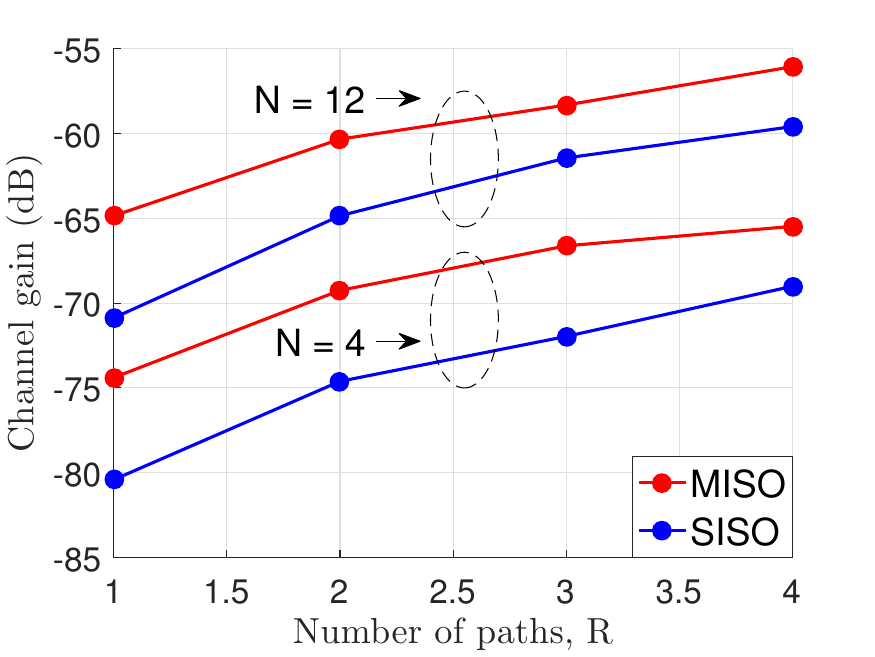}
    \caption{Channel gain versus the number of paths $R$.}
    \label{fig_4}
    \vspace{-0.3cm}
\end{figure}

In Fig. \ref{fig_4}, we examine the channel gain versus the number $R$ of propagation paths, with other system parameters remaining unchanged. It is observed that the channel gain increases as $R$ increases, which indicates that multipath propagation is beneficial for FIM-assisted wireless communication systems. This is because as a larger number of propagation paths are involved, it is more likely to make the signal copies from these different paths add constructively by morphing the 3D surface shape of the FIM.

\section{Conclusion}
FIM introduces unprecedented flexibility compared to traditional rigid RIS. By morphing its surface shape, FIM gains extra design DoF to manipulate the propagation environment, thus further improving the channel gain. In the SISO scenario, we maximized the FIM-aided channel gain by utilizing PSO and MIGD to optimize the FIM surface shape and phase shifts. In the MISO scenario, we proposed an efficient alternating optimization algorithm to optimize the FIM surface shape, the phase shifts, and the transmit beamforming for maximizing the channel gain. Our numerical results demonstrate that the FIM yields over $3$ dB improvement compared to traditional rigid RIS under all scenarios considered, by leveraging its surface shape morphing capability. 

Nonetheless, robust surface shape morphing algorithm design and performance evaluation of FIM-assisted systems in more complex networks deserve further investigation. Given that FIM introduces mechanical surface shape morphing, it requires further investigation to elucidate the tradeoff between performance gains and power consumption.
\bibliographystyle{IEEEtran}
\bibliography{reference}
\end{document}